\documentclass[prd,twocolumn,nofootinbib,aps,tightenlines,preprintnumbers,notitlepage,longbibliography,superscriptaddress]{revtex4-1}

\usepackage{amssymb}
\usepackage{amsmath}
\usepackage{graphicx}
\usepackage{hyperref}
\usepackage{natbib}
\usepackage{color}

\begin{document}

\title{Reheating constraints to WIMPflation}

\author{Lingyuan Ji}
\email{lingyuan.ji@jhu.edu}
\author{Marc Kamionkowski}
\email{kamion@jhu.edu}
\affiliation{Department of Physics and Astronomy, Johns Hopkins University, 3400 North Charles Street, Baltimore, Maryland 21218, USA}

\begin{abstract}
Analyses of inflation models are usually conducted assuming a specific range---e.g., $N_k \simeq 50-60$--of the number $N_k$ of $e$-folds of inflation. However, the analysis can also be performed by taking into account constraints imposed by the physics of reheating.  In this paper, we apply this analysis to a class of ``WIMPflation'' models in which the inflaton also plays the role of dark matter. Our analysis also updates prior WIMPflation work with more recent Planck 2018 data. With this new analysis, inflaton potentials $V(\phi)=\lambda\phi^4$ and $\lambda \phi_0^4[1-\cos(\phi/\phi_0)]^2$ are ruled out, while $V(\phi)=\lambda \phi_0^4\{1-\exp[-(\phi/\phi_0)^2]\}^2$ is slightly disfavored, and $V(\phi)=\lambda\phi_0^4\tanh^4(\phi/\phi_0)$ is only viable for certain reheating conditions.  In addition, we also discuss for the first time the effect of post-reheating entropy production (from, e.g., cosmological phase transitions) in this reheating-physics analysis. When accounted for, it decreases the number of $e$-folds through $\Delta N_k=-(1/3)\ln(1+\gamma)$, where $\gamma\equiv\delta s/s$ is the fractional increase in entropy. We discuss briefly the possible impact of entropy production to inflation-model constraints in earlier work.

\end{abstract}

\maketitle

\section{Introduction}

Cosmic inflation driven by a slowly rolling scalar field has been conjectured as the standard solution to many problems in cosmology \cite{Guth:1980zm,Linde:1981mu,Albrecht:1982wi}. During inflation, the inflaton rolls down a sufficiently flat potential, providing the energy density to inflate the Universe, as well as generating primordial perturbations. If the inflaton is stable, it may also constitute the dark matter, a possibility first proposed in Refs.\ \cite{Kofman:1994rk, Kofman:1997yn} and elaborated most recently in Ref.\ \cite{Hooper:2018buz}. In this scenario, reheating begins (see Ref.\ \cite{Allahverdi:2010xz} for a review) after inflation ends, but before the inflaton reaches the minimum of its potential.  Quanta that arise as oscillations about the inflaton-potential minimum then provide a dark-matter candidate.

In several previous papers, it has been shown that the number of $e$-folds needed for inflation is related to the physics of reheating. The basic idea was first discussed in Ref.\ \cite{Dodelson:2003vq} with instantaneous reheating; it was later extended to reheating with a constant equation-of-state parameter \cite{Liddle:2003as, Martin:2010kz, Adshead:2010mc}. Refs.\ \cite{Dai:2014jja, Munoz:2014eqa,Cook:2015vqa} then applied this method to different inflation models using post-Planck data.

In this paper, we use this method to analyze a new set of WIMPflation models \cite{Hooper:2018buz}; we also update earlier work through the inclusion of more recent Planck 2018 data. Furthermore, we discuss for the first time the effect of post-reheating entropy production on the results of the analysis. If reheating happens at a high energy scale, then any of a number of post-reheating events may have led to entropy production. Such entropy production is even conceivable at the electroweak scale \cite{Chaudhuri:2017icn}.

The structure of this paper is as follows. In Section~\ref{sec:models-inflation} and Section~\ref{sec:models-reheating}, the models of WIMPflation and reheating are introduced respectively. In Section~\ref{sec:entropy-production}, we discuss the effects of entropy production. In Section~\ref{sec:results}, results for the WIMPflation models we consider here are presented, and in Section~\ref{sec:conclusions} we make concluding remarks.

\section{Models}
\label{sec:models}

\subsection{Inflation}
\label{sec:models-inflation}

We consider a class of ``WIMPflation'' models \cite{Hooper:2018buz} in which a single field $\phi$ acts as the inflaton and as dark matter. The Lagrangian for the inflaton is
\begin{equation}
	\mathcal{L}=\frac{1}{2}(\partial^{\mu}\phi)(\partial_{\mu}\phi)-V(\phi)+\mathcal{L}_\mathrm{int},
\end{equation}
where $\mathcal{L}_\mathrm{int}$ contains the interaction terms required to reheat the Universe. The inflaton potential is
\begin{equation}
	V(\phi)=\frac{1}{2}m_\phi^2\phi^2+\lambda\phi_0^4f\left(\frac{\phi}{\phi_0}\right),
\end{equation}
where the first term gives the inflaton mass $m_\phi$; $f(\phi/\phi_0)$ is a function with a vanishing second derivative $f''(0)=0$; $\phi_0$ is a constant with mass dimension 1, determining the scale of inflaton field. Here, $\lambda$ is a dimensionless constant which together with $\phi_0$ fixes the energy scale of inflation. Here we consider the following functional forms \cite{Hooper:2018buz}:
\begin{equation}
	f(x=\phi/\phi_0)=
	\begin{cases}
		x^{4},\\
		\arctan^{4}x,\\
		\tanh^{4}x,\\
		[1-\exp(-x^{2})]^{2},\\
		(1-\cos x)^{2},\\
		x^{4}/(1+x^{2})^{2}.
	\end{cases}
\end{equation}

In this work, the dynamics of inflation are considered within the conventional slow-roll regime \cite{Baumann:2009ds} and are described by slow-roll parameters,
\begin{align}
	\epsilon &\equiv \frac{M_P^2}{2}\left(\frac{V'}{V}\right)^2 = \frac{2M_P^2(\lambda \phi_0^3 f'+\phi m_\phi^2)^2}{(2\lambda \phi_0^4 f+\phi^2 m_\phi^2)^2}, \\
	\eta &\equiv M_P^2\frac{V''}{V} = \frac{2M_P^2(\lambda\phi_{0}^{2}f''+m_{\phi}^{2})}{2\lambda\phi_{0}^{4}f+\phi^{2}m_{\phi}^{2}},
\end{align}
where $M_P=2.435\times 10^{18}\,\mathrm{GeV}$ is the reduced Planck mass. Inflation ends when $\epsilon(\phi_\mathrm{end})=1$. We denote all quantities evaluated when a specific $k$-mode exits the horizon (i.e.\ $k=a_kH_k$) with subscript $k$. Therefore, the number of $e$-folds of inflation is given by the integral
\begin{equation}
	N_k=\int^{t_\mathrm{end}}_{t_k} Hdt\approx \int_{\phi_\mathrm{end}}^{\phi_k}\frac{V}{V'}\frac{d\phi}{M_P^2},
\end{equation}
whose exact form depends on the parametrization of $f$. In the last step, we used the slow-roll approximation. Furthermore, the primordial amplitude $A_s$, tensor-to-scalar ratio $r$ and the primordial tilt $n_s$ can be expressed using the slow-roll parameters as
\begin{equation}\label{eq:observables}
	A_{s}=\frac{1}{24\pi^2\epsilon}\frac{V}{M_P^4},\quad r=16\epsilon,\quad n_s=1-6\epsilon+2\eta.
\end{equation}
These quantities should all be evaluated at the scale $k$ of interest. The subscripts $k$ here are suppressed for simplicity.

The inflation models we discussed here involve 3 parameters $(m_\phi,\phi_0,\lambda)$. In principle, with precise knowledge of the 3 observables mentioned in Eq.\ (\ref{eq:observables}), all parameters of the model can be derived for a specific $f$. However, in practice only $A_s$ and $n_s$ are relatively well determined. We therefore in this work follow standard practice in fixing $A_s$ and then presenting predictions in the remaining $n_s$-$r$ parameter space.  We have verified that our results are insensitive to changes to $A_s$ that are within its $3\sigma$ range.

\subsection{Reheating}
\label{sec:models-reheating}

\begin{figure}
	\includegraphics[width=0.5\textwidth]{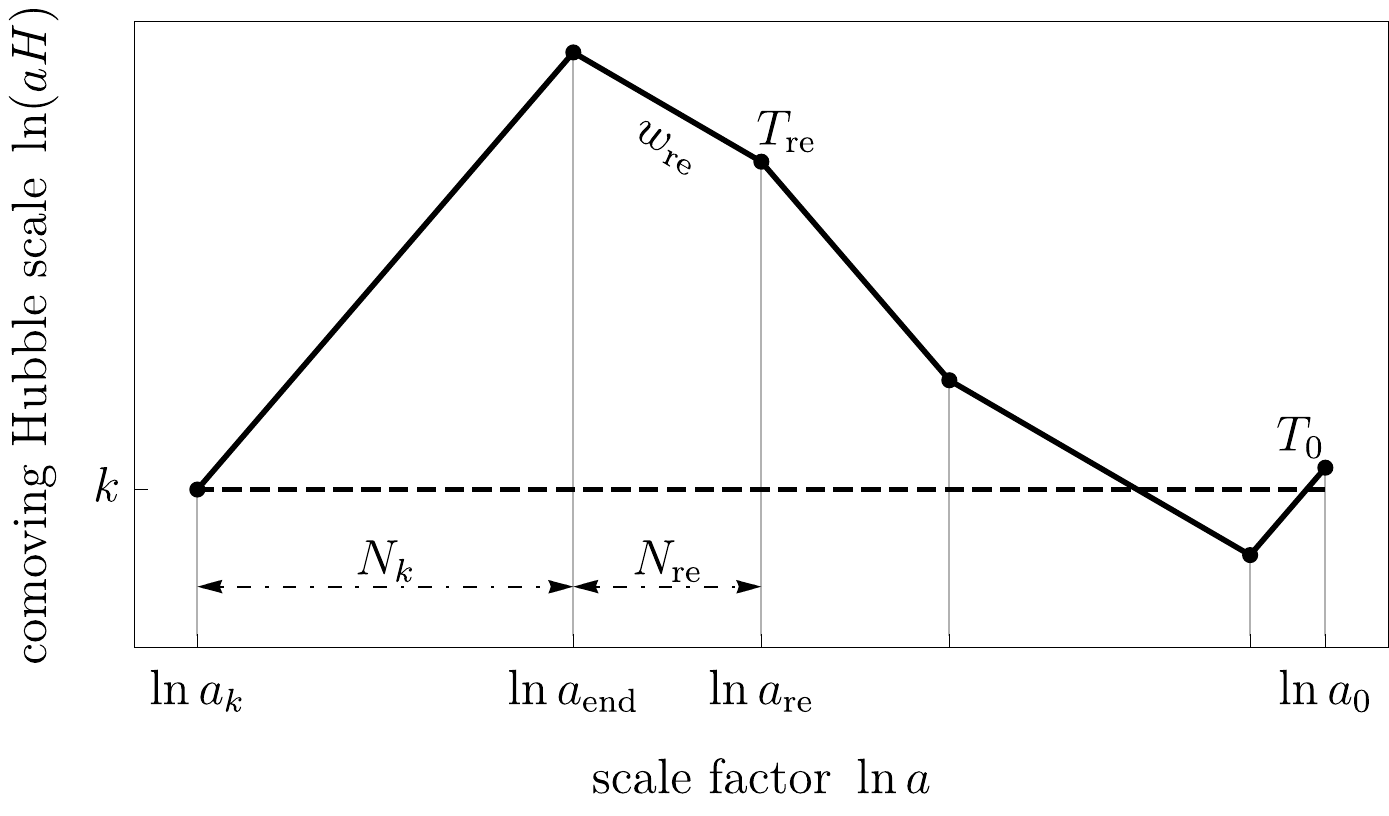}
	\caption{Sketch of the comoving Hubble scale $\ln(aH)$ as a function of the scale factor $\ln a$. Here, $a_k$, $a_\mathrm{end}$, $a_\mathrm{re}$, and $a_0$ are the scale factor when $k$-mode exits the horizon, end of inflation, end of reheating, and today, respectively. Here $N_k$ and $N_\mathrm{re}$ are the number of $e$-folds of inflation and reheating respectively. Also, $T_\mathrm{re}$ is the temperature at the end of reheating. The slope of the line during reheating is dictated by the reheating equation-of-state parameter $w_\mathrm{re}$.}\label{fig:hubble}
\end{figure}

Here we discuss, following Refs.\ \cite{Dai:2014jja, Munoz:2014eqa}, how to calculate the number $N_\mathrm{re}$ of $e$-folds of expansion during reheating and also the reheat temperature $T_\mathrm{re}$, parametrizing the reheating epoch by a constant equation-of-state parameter $w_\mathrm{re}$. For reference, we sketch in Fig.\ \ref{fig:hubble} the comoving Hubble scale as a function of scale factor and define there some relevant quantities.

For a constant equation-of-state parameter expansion, the conservation of energy momentum relates energy density $\rho$ to scale factor $a$ as $\rho\propto a^{[-3(1+w_\mathrm{re})]}$. Therefore the number of $e$-folds of expansion during reheating can be written as
\begin{equation}\label{eq:n-re}
	N_\mathrm{re}=\frac{1}{3(1+w_\mathrm{re})}\ln\frac{\rho_\mathrm{end}}{\rho_\mathrm{re}},
\end{equation}
where $\rho_\mathrm{end}$ and $\rho_\mathrm{re}$ are the energy density of the Universe at the end of inflation and at the end of reheating respectively. Also, $\rho_\mathrm{end}=(1+\kappa)V_\mathrm{end}$, where $\kappa=(3/\epsilon-1)^{-1}$ is the ratio of kinetic energy to potential energy during inflation and $V_\mathrm{end} \equiv V(\phi_\mathrm{end})$. At the end of inflation, $\epsilon\approx 1$, so $\kappa\approx 1/2$, and $\rho_\mathrm{re}$ can be expressed as $(\pi^2/30)g_\mathrm{re}T_\mathrm{re}^4$ with $g_\mathrm{re}$ the effective number of relativistic degrees of freedom for energy at full thermalization. The reheating temperature $T_\mathrm{re}$ can be related to the CMB temperature $T_0$ today via entropy conservation,
\begin{equation}\label{eq:entropy}
	a_\mathrm{re}^3 g_{s,\mathrm{re}}T_\mathrm{re}^3=a_0^3\left(2T_0^3+\frac{7}{8}\times 6T_{\nu0}^3\right),
\end{equation}
where $g_{s,\mathrm{re}}$ is the effective number of relativistic degrees of freedom for entropy at the end of reheating. Combining Eq.\ (\ref{eq:entropy}) with the current neutrino temperature $T_{\nu0}=(4/11)^{1/3}T_0$, we arrive at the relation
\begin{equation}\label{eq:t-re}
	\frac{T_\mathrm{re}}{T_0}=\left(\frac{43}{11g_{s,\mathrm{re}}}\right)^{1/3}\frac{a_0}{a_\mathrm{re}}.
\end{equation}

Eqs.\ (\ref{eq:n-re}) and (\ref{eq:t-re}) are not sufficient to determine $T_\mathrm{re}$ and $N_\mathrm{re}$. Information on how much the Universe has expanded since the end of reheating is needed. The equations can be closed with the relation,
\begin{equation}\label{eq:geometric-relation}
	\frac{k}{a_{0}H_{0}}=\frac{a_{k}}{a_{\text{end}}} \frac{a_{\text{end}}}{a_{\text{re}}} \frac{a_{\text{re}}}{a_{0}} \frac{H_{k}}{H_{0}},
\end{equation}
where $H_k^2=V(\phi_k)/(3M_P^2)$ and $H_0^2=\rho_\mathrm{crit}/(3M_P^2)$, with $\rho_\mathrm{crit}$ being the critical density. Solving Eqs.\ (\ref{eq:n-re}), (\ref{eq:t-re}), and (\ref{eq:geometric-relation}), we have
\begin{align}
	N_{\text{re}}=&\frac{1}{1-3w_{\text{re}}}\bigg[\ln\frac{T_{0}^{4}}{(1+\kappa)V_{\text{end}}}-4N_{k}-4\ln\frac{k}{a_{0}H_{0}}\notag\\
	&+2\ln\frac{\rho_{k}}{\rho_{\text{crit}}}+\frac{4}{3}\ln\frac{43}{11g_{s,\text{re}}}+\ln\frac{\pi^{2}g_{\text{re}}}{30}\bigg],\label{eq:n-re-sol}\\
T_{\text{re}}=&\exp\left[-\frac{3}{4}(1+w_{\text{re}})N_{\text{re}}\right]\left(\frac{30}{g_{\text{re}}\pi^{2}}\right)^{\frac{1}{4}}(1+\kappa)^{\frac{1}{4}}V_{\text{end}}^{\frac{1}{4}}\label{eq:t-re-sol}.
\end{align}

Reheating requires $N_\mathrm{re}\geq 0$, and we refer to the case where $N_\mathrm{re}=0$ as instantaneous reheating. The conditions for the Universe to thermalize before BBN or the EW phase transition are $T_\mathrm{re}>T_\mathrm{BBN}$ and $T_\mathrm{re}>T_\mathrm{EW}$, respectively. The agreement between the observed light-element abundances and those predicted by BBN precludes the possibility of significant entropy production (to be discusses in the next Section) unless $T_\mathrm{re}>T_\mathrm{BBN}$.

\section{Entropy production}
\label{sec:entropy-production}

In prior related work \cite{Dai:2014jja, Munoz:2014eqa}, it is assumed that there is no significant entropy production after reheating, but there are many ways in which this assumption might be invalidated, through, for example, out-of-equilibrium particle decays or any of a number of phase transitions that could conceivably occur after reheating but before BBN. Even in the Standard Model, there is a small amount of entropy production \cite{Chaudhuri:2017icn}. Fortunately the effects of reheating on the analysis are, as we now show, easy to take into account.

Suppose there is a fractional increase $\gamma \equiv \delta s/s>0$ in the entropy density between reheating and BBN. If so, Eq.\ (\ref{eq:entropy}) is augmented to
\begin{equation}
		a_\mathrm{re}^3 g_{s,\mathrm{re}}T_\mathrm{re}^3(1+\gamma) = a_0^3\left(2T_0^3+\frac{7}{8}\times 6T_{\nu0}^3\right).
\end{equation}
The effects of nonzero $\gamma$ on the results of Section\ \ref{sec:models-reheating} can therefore be taken into account simply with the replacement $g_\mathrm{s,re} \rightarrow (1+\gamma)g_\mathrm{s,re}$ in Eqs.\ (\ref{eq:n-re-sol}) and (\ref{eq:t-re-sol}). By keeping $N_\mathrm{re}$ unchanged in Eq.\ (\ref{eq:n-re-sol}), we infer that the number $N_k(\gamma)$ of $e$-folds of inflation between the time a distance scale $k$ exits the horizon and the end of inflation becomes
\begin{equation}
	N_k(\gamma) = N_k(\gamma=0) + \frac{1}{3} \ln \frac{1}{1+\gamma}.
\end{equation}
The Standard Model expectation for $\gamma$ from the EW scale to BBN scale is $\gamma=0.13$ \cite{Chaudhuri:2017icn}, which corresponds to only a negligible change $\Delta N_k \equiv N_k(\gamma) - N_k(\gamma=0) \simeq -0.04$. Supercooling during a post-reheating first-order phase transition (essentially, a later short period of inflation \cite{Silk:1986vc}) might conceivably increase $\gamma$ by several orders of magnitude.

The possibility of significant post-reheating entropy production has implications for the results of prior work in which entropy conservation was assumed.  For example, the minimum tensor-to-scalar ratios inferred in Ref.~\cite{Munoz:2014eqa} will increase (as the curves in their Fig.~4 move to the left).  The $\alpha=2/3$ and $\alpha=1$ models which were ruled out in Ref.~\cite{Dai:2014jja} may be revived, as the curves in their Fig.~2 will move to the left, as $n_s$ is a monotonically increasing function of $N_k$ in those models. Still, any such changes, if they are to be consequential, would require $\gamma \sim e^{3\times 5} \sim 3\times 10^6$; i.e., a fairly radical augmentation of the post-reheating expansion history.

\section{Results for WIMPflation models}
\label{sec:results}

The model space (inflation and reheating) is characterized by 5 parameters
\begin{equation}
	\{m_\phi,\phi_0,\lambda,\phi_k;w_\mathrm{re}\}.
\end{equation}
Here the first 3 parameters are those for the inflation model described in Section \ref{sec:models-inflation}. The 4th parameter $\phi_k$ is the inflaton field value when the $k$-mode exits the horizon. It dictates the length of inflation. The last one is the reheating equation-of-state parameter described in Section \ref{sec:models-reheating}. Given a specific model, the following 6 quantities can be calculated
\begin{equation}
	\{N_k,A_s,n_s,r;T_\mathrm{re},N_\mathrm{re}\},
\end{equation}
using the formulas derived in Section \ref{sec:models-inflation} and \ref{sec:models-reheating}. We are interested in the predictions of the model in the $(n_s,r)$ plane with some fixed values of $T_\mathrm{re}$ or $N_\mathrm{re}$. In this case, the value of $\phi_k$ is inferred from those fixed values.

To simplify the analysis, we fix some parameters to their canonical values: $g_\mathrm{re}=g_\mathrm{s,re}=100$ (roughly the number of degrees of freedom in the Standard Model), $T_0=9.7\times 10^{-32}\,M_P,\ \rho_\mathrm{crit}=1.2\times 10^{-120}\,M_P^{4}$. We fix the inflaton mass $m_\phi=10\,\mathrm{MeV}$, but the results we present are unchanged even if it is as big as 100 TeV.  Here we assume entropy conservation (i.e.\ $\gamma=0$) and then discuss below how significant entropy production affects the results. We also set the pivot scale\footnote{The Planck 2018 pivot scale is $0.05\,\mathrm{Mpc}^{-1}$ \cite{Akrami:2018odb}. However, the tensor-to-scalar ratio $r$ is also quoted at $0.002\,\mathrm{Mpc}^{-1}$ to facilitate comparison with earlier Planck constraints. Here we use $r$ measured at $k=0.002\,\mathrm{Mpc}^{-1}$.} to $k=0.002\,\mathrm{Mpc}^{-1}=8\, a_0H_0$. Moreover, as stated in Section~\ref{sec:models-inflation}, the Planck 2018 best fit for $\ln(10^{10}A_s)|_{k=0.05\,\mathrm{Mpc}^{-1}}=3.043$ is used to infer the value of $\lambda$ every time we move to a new point in model space. Note that we need to convert this to our pivot scale, which gives $\ln(10^{10}A_s)|_{k=0.002\,\mathrm{Mpc}^{-1}}=3.155$.

\begin{figure*}
	\includegraphics[width=0.49\textwidth]{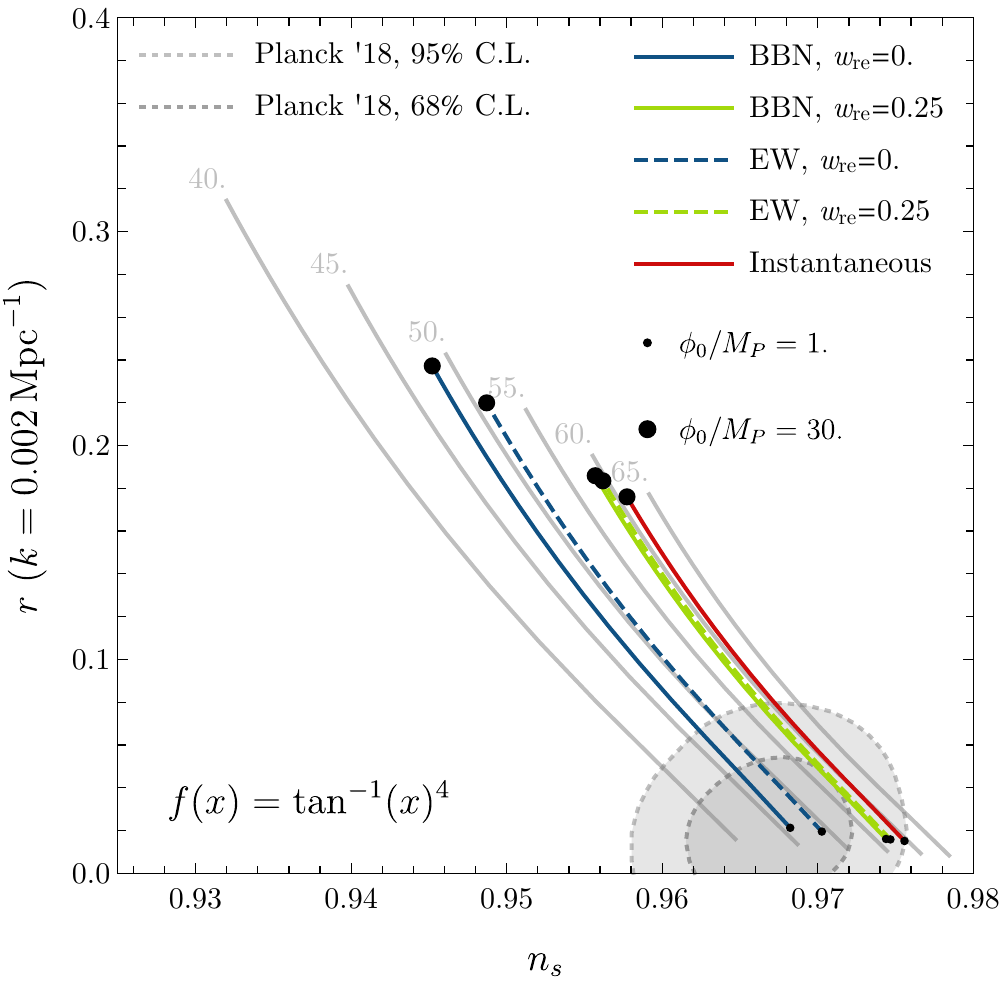}
	\includegraphics[width=0.49\textwidth]{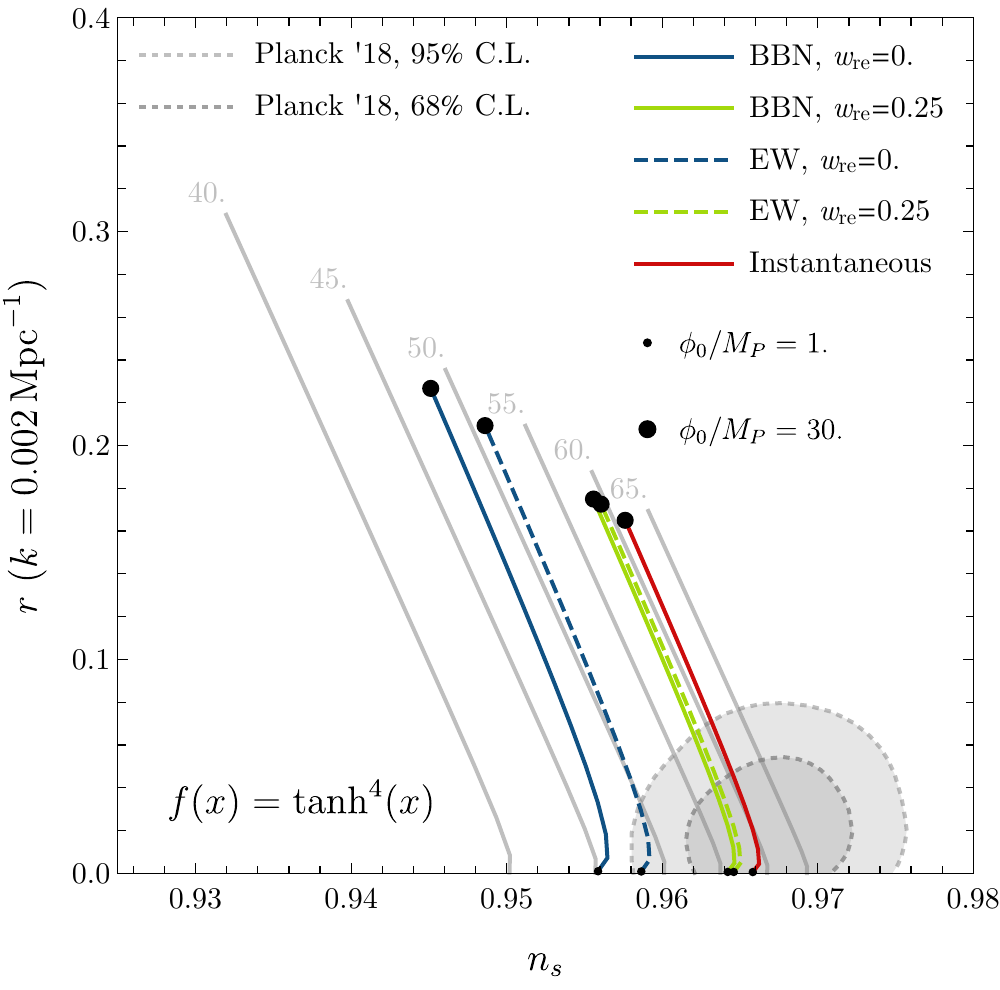}\\
	\includegraphics[width=0.49\textwidth]{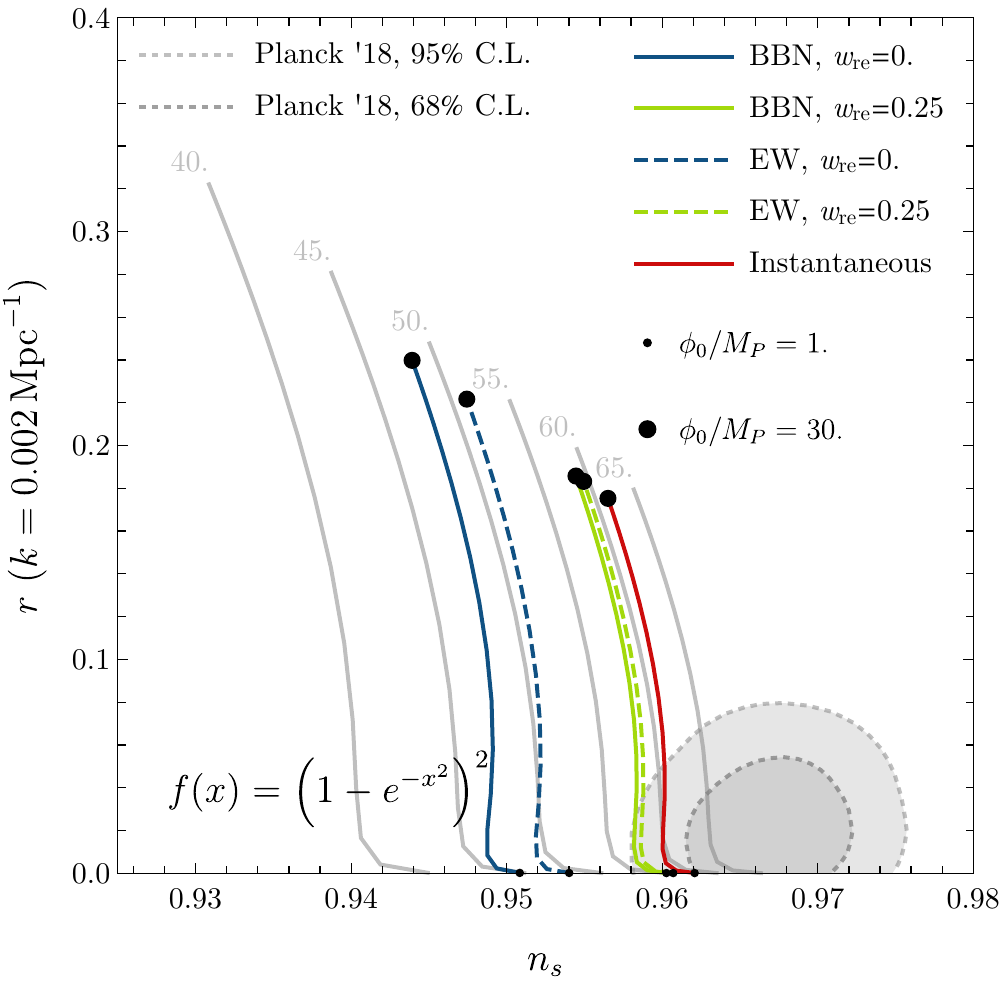}
	\includegraphics[width=0.49\textwidth]{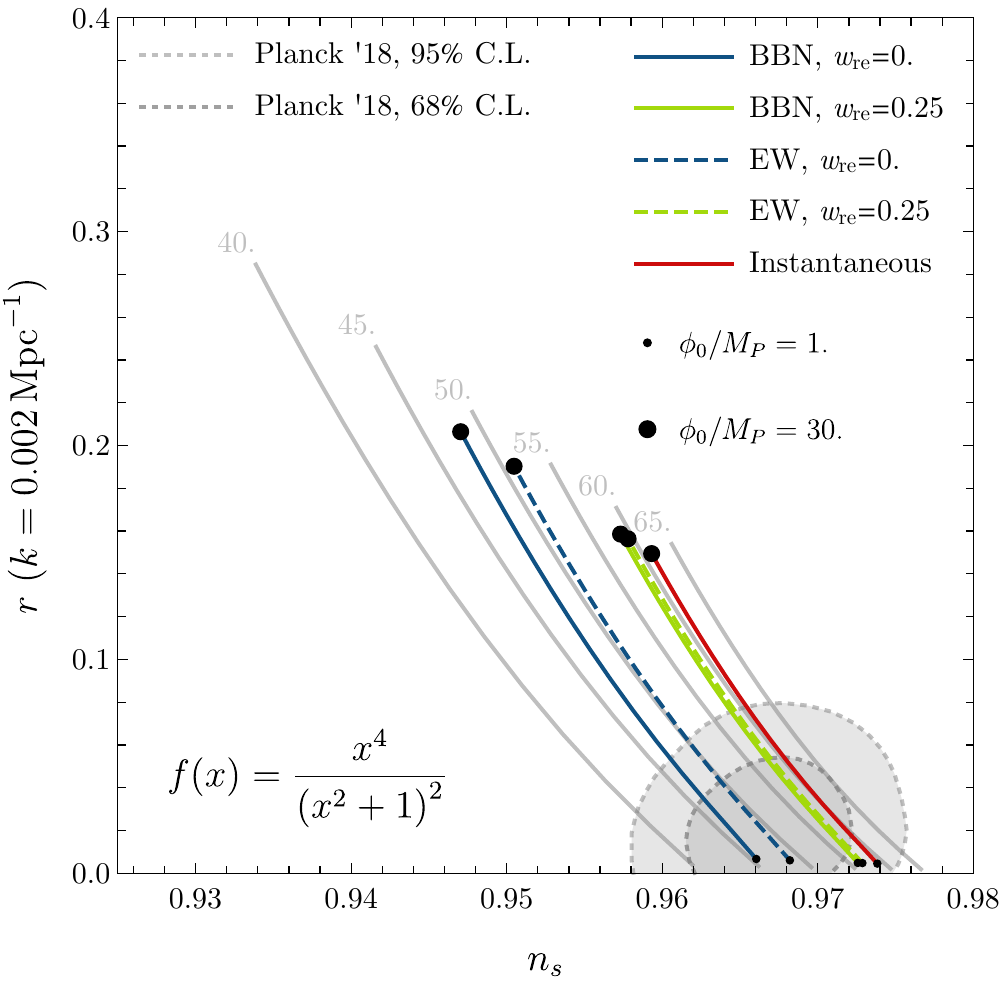}
	\caption{The $n_s$-$r$ plot for different models. The 4 panels correspond to 4 different inflaton potentials. In each panel, the solid blue (green) line represents a BBN scale reheating with equation-of-state parameter $w_\mathrm{re}=0\ (0.25)$, and the dashed blue (green) line represents an EW scale reheating with equation-of-state parameter $w_\mathrm{re}=0\ (0.25)$. The solid red line represents an instantaneous reheating. Along each line, the parameter $\phi_0$ varies from $1\,M_P$ to $30\,M_P$. The set of solid gray lines in the background benchmarks the number $N_k$ of $e$-folds of different scenarios. Fig.\ 2 in Ref.\ \cite{Hooper:2018buz} showed two of them for $N_k=50\ \mathrm{and}\ 60$. The posterior contours are extracted from the Planck 2018 results for inflation. The choices of all unmentioned parameters in the plot are specified in Section~\ref{sec:results}. The model $f(x)=x^4$ is not shown as $\lambda \phi_0^4 f(\phi/\phi_0)=\lambda \phi^4$ turns out to be independent of $\phi_0$. Nevertheless, it is the large $\phi_0$ limit for all other models. The model $f(x)=(1-\cos x)^2$ is omitted since even the largest allowed region (between solid blue and red lines) is excluded by Planck data.}\label{fig:ns-r}
\end{figure*}

The results of the analysis are shown in the $(n_s,r)$ plane in Fig.\ \ref{fig:ns-r}. Predictions with different reheating conditions are plotted for various inflation models. Predictions obtained assuming \emph{a priori} some fixed number $N_k$ of $e$-folds are plotted as lighter curves for comparison. As illustrated in the plot, instantaneous reheating implies that $N_k\sim 60$ (depending in detail on the model), regardless of the equation-of-state parameter. This serves as an upper bound on the number of $e$-folds for any of this class of models. For EW scale reheating, our analysis yields $N_k\sim 57\ (50)$ (depending in detail on the model) for $w_\mathrm{re}=0.25\ (0)$; for BBN scale reheating, our analysis yields $N_k\sim 56\ (46)$ (depending in detail on the model) for $w_\mathrm{re}=0.25\ (0)$. Although the curve with a specific reheating temperature $T_\mathrm{re}$ is very close to a curve with some constant number $N_k$ of $e$-folds, they do not exactly overlap. At a specific reheating scale, a model with a larger $\phi_0$ parameter, or, equivalently, a larger tensor-to-scalar ratio $r$, tends to involve a longer period of inflation.

The regions between the EW (BBN) lines and the instantaneous lines are those allowed by the requirement that the reheat temperature be higher than the EW (BBN) scale. They can be compared to the Planck measurement (Fig.~\ref{fig:ns-r}, gray blob) to rule out certain inflation models. Such a comparison completely excludes $f(x)=x^4$ and $(1-\cos x)^2$, as even the largest allowed region is incompatible with Planck measurement. It also slightly disfavors $f(x)=[1-\exp(-x^2)]^2$, but still leaves an opening at scenarios close to instantaneous reheating. For $f(x)=\tanh^4 x$, if $w_\mathrm{re}=0$, only EW scale reheating can comply with the data. All other models are left untouched.

If entropy production exists, according to Section\ \ref{sec:entropy-production}, $N_k$ decreases. All curves presented in Fig.\ \ref{fig:ns-r} will move to the left to a lower number of inflation $e$-folds.

\section{Conclusions}
\label{sec:conclusions}

In this paper, we have studied the implications of reheating on WIMPflation models. We have parametrized uncertainties in reheating physics in terms of a constant equation-of-state parameter. We then explore the allowed parameter space of several WIMPflation models with this parametrization, including the most recent CMB measurements from Planck.

\begin{table}
  \begin{tabular}{|c|cc|}\hline
    $(N_k^\mathrm{min},N_k^\mathrm{max})$ & $w_\mathrm{re}=0$ & $w_\mathrm{re}=0.25$ \\
    \hline
    $T_\mathrm{re}=T_\mathrm{EW}$ & $(50,60)$ & $(57,60)$ \\
    $T_\mathrm{re}=T_\mathrm{BBN}$ & $(46,60)$ & $(56,60)$ \\ \hline
  \end{tabular}
  \caption{The approximately allowed range of e-folds $N_k$ dictated by the scale  $T_\mathrm{re}$ of reheating and the equation-of-state parameter $w_\mathrm{re}$. $N_k^\mathrm{max}\sim 60$ is determined by instantaneous reheating, thus irrelevant to $T_\mathrm{re}$ and $w_\mathrm{re}$. The exact shapes of the constraints are different from a constant number of $e$-folds range and depend on the detailed shape of the inflaton potential. Refer to Figure~\ref{fig:ns-r} for more information.}\label{tab:e-folds}
\end{table}

We discover that the allowed ranges of inflation observables is very similar to those allowed by a given range of the number $N_k \in [N_k^\mathrm{min},N_k^\mathrm{max}]$ of $e$-folds Here, however, $(N_k^\mathrm{min},N_k^\mathrm{max})$ are not of the canonically taken value $(50,60)$, but related to the scale $T_\mathrm{re}$ of reheating and the equation-of-state parameter $w_\mathrm{re}$. For the models we investigate, this relation can be summarized approximately in Table~\ref{tab:e-folds}. The constraints are provided more precisely in Fig.~\ref{fig:ns-r}. By comparing the model predictions to constraints from Planck 2018, the data completely exclude $f(x)=x^4$ and $ f(x) =(1-\cos x)^2$, and slightly disfavor $f(x)=[1-\exp(-x^2)]^2$. The $f(x)=\tanh^4 x$ model survives for some range of reheating parameters, and $f(x)=\arctan^4 x$ and $f(x)=x^4/(1+x^2)^2$ remain viable.

We also show how post-reheating entropy production will affect the results of the analysis as well as results in related earlier work. When taken into account, it will decrease the number of $e$-folds of inflation through $\Delta N_k=-(1/3)\ln(1+\gamma)$, with $\gamma \equiv\delta s/s$ being the fractional increase in entropy.  However, for the Standard Model EW phase transition with only one Higgs, the effect is negligible. In some extended models with multiple Higgs fields, the production may be significant.  We also discussed the implications of entropy production on the results of some prior papers.

\begin{acknowledgements}
We thank K.\ Boddy and T.\ Tenkanen for useful discussions. This work was supported by NSF Grant No.\ 1519353, NASA NNX17AK38G, and the Simons Foundation.
\end{acknowledgements}

\end{document}